\documentclass[aps,twocolumn,superscriptaddress,nofootinbib,preprintnumbers]{revtex4}

\usepackage[dvips]{color}
\usepackage[normalem]{ulem}
\usepackage{amsmath}
\usepackage{enumerate}
\usepackage{amsfonts}
\usepackage{epsfig}
\usepackage{yfonts}

\usepackage{graphicx}
\usepackage{bm}

\def\eg{{\it e.g. }}
\def\ie{{\it i.e.}}

\def\({\left(}
\def\){\right)}
\def\[{\left[}
\def\]{\right]}
\def\<{\langle}
\def\>{\rangle}

\def\CC{{\cal C}}

\def\CO{{\cal O}}

\def\Dslash{\rlap{\hskip0.2em/}D}

\def\tr{\mathop{\rm tr}}

\newcommand\half{{\ensuremath{\frac{1}{2}}}}

\newcommand\vev[1]{{\ensuremath{\left\langle{#1}\right\rangle}}}

\newcommand{\be}{\begin{equation}}
\newcommand{\ee}{\end{equation}}
\newcommand{\bea}{\begin{eqnarray}}
\newcommand{\eea}{\end{eqnarray}}
\newcommand{\bwt}{\begin{widetext}}
\newcommand{\ewt}{\end{widetext}}

\newcommand{\bi}{\begin{itemize}}
\newcommand{\ei}{\end{itemize}}
\newcommand{\ben}{\begin{enumerate}}
\newcommand{\een}{\end{enumerate}}
\newcommand{\bca}{\begin{cases}}
\newcommand{\eca}{\end{cases}}
\newcommand{\bln}{\begin{align}}
\newcommand{\eln}{\end{align}}
\newcommand{\bst}{\begin{split}}
\newcommand{\est}{\end{split}}

\newcommand\sG{{\ensuremath{{\mathcal G}}}}

\renewcommand{\Im}{\textrm{Im}\,}
\renewcommand{\Re}{\textrm{Re}\,}

\newcommand{\ut}{\underline t}
\newcommand{\umu}{\underline \mu}
\newcommand{\ux}{\underline x}
\newcommand{\uy}{\underline y}

\newcommand{\ur}{\underline r}

\newcommand{\cF}{\mathcal{F}}
\newcommand{\cO}{\mathcal{O}}
\newcommand{\bI}{\mathbf{I}}
\newcommand{\bII}{\mathbf{II}}

\def\Dslash{\rlap{\hskip0.2em/}D}

\def\a{\alpha}

\def\cF{{\mathcal F}}
\def\cIR{c_{{\rm IR}}}
\def\scalar{\varphi}
\def\spinor{\zeta}
\def\mscalar{m_\varphi}
\def\mspinor{m_\zeta}
\def\qscalar{q_\varphi}
\def\qspinor{q_\zeta}

\begin{document}

\title{Photoemission ``experiments" on holographic superconductors}

\preprint{MIT-CTP/4088, NSF-KITP-09-150}

\author{Thomas Faulkner}
\affiliation{KITP, Santa Barbara, CA 93106}

\author{Gary T. Horowitz}
\affiliation{Department of Physics, UCSB, Santa Barbara, CA 93106}

\author{John McGreevy}
\affiliation{Center for Theoretical Physics, \\
Massachusetts Institute of Technology,
Cambridge, MA 02139 }

\author{Matthew M. Roberts}
\affiliation{Department of Physics, UCSB, Santa Barbara, CA 93106}

\author{David Vegh}
\affiliation{Simons Center for Geometry and Physics, Stony Brook University, Stony Brook, NY 11794-3636}
\affiliation{Center for Theoretical Physics, \\
Massachusetts Institute of Technology,
Cambridge, MA 02139 }
%
\begin{abstract}

We study the effects of a
superconducting condensate
on holographic Fermi surfaces. With a suitable coupling between the fermion and the condensate, there are stable quasiparticles with a gap.
We find some similarities with the
phenomenology of the cuprates:
in systems whose normal state
is a non-Fermi liquid with
no stable quasiparticles,
a stable quasiparticle peak appears
in the condensed phase.

\end{abstract}

\today

\maketitle

\tableofcontents

\section{Introduction}

The problem of what happens when a large number of interacting fermions
get together remains interesting despite many decades of work.
The sign problem obstructs a numerical solution,
leaving us to do experiments or theorize.
The metallic states of such systems that
are well-understood theoretically are Fermi liquids.
The basic assumption of this theory is that
the states of the interacting system
can be usefully put in correspondence with those
of a collection of the same number of free fermions;
in particular this means that the low-lying excitations
of the system are long-lived quasiparticles.

This assumption fails in many strongly-correlated materials.
Quite a bit of effort has been made to understand what
replaces the Fermi liquid theory in the absence of stable quasiparticles
\cite{Holstein:1973zz,reizer,baym,Polchinski:1993ii,Nayak:1993uh,Halperin:1992mh,altshuler:1994,Schafer:2004zf,Boyanovsky,sungsikgaugefield,nagaosa,Patrick,fradkin,nave}.
We believe that it is fair to say that it would be valuable
to have a non-perturbative description of such a state of matter.
Inspired by work of Sung-Sik Lee \cite{Lee:2008xf},
a class of non-Fermi liquids
was recently found
\cite{Liu:2009dm, Faulkner:2009wj} (see also \cite{Cubrovic:2009ye, soojong})
using holographic duality \cite{AdS/CFT}.
This allows us to study observables
of the strongly-coupled system using simple
gravity calculations.
For a review of these techniques in the present context, see
\cite{Sachdev:2008ba, Hartnoll:2009sz, Herzog:2009xv, McGreevy:2009xe}.

The analysis of \cite{Liu:2009dm, Faulkner:2009wj}
applied to CFTs with a gravity dual, a conserved $U(1)$ current,
and a charged fermionic operator.
Depending on the charge and dimension of the operator,
it is possible to find Fermi liquid behavior,
in the sense that the spectral function exhibits
stable quasiparticles,
or non-Fermi liquid behavior.
At the boundary between these behaviors,
one finds a marginal Fermi liquid,
which arises as a phenomenological model \cite{varma} of the
strange metal phase of the cuprate superconductors
(the resistive state at temperatures larger than the critical temperature $T_c$
for superconductivity, at a doping level which maximizes $T_c$).
In this case, the contribution of such a Fermi surface
to the resistivity also
has the linear temperature dependence
observed in the strange metal
\cite{resistivitypaper}.

The calculation of the fermion spectral functions
was done by solving the Dirac equation in a charged black hole background.
The extremal Anti-de Sitter (AdS) Reissner-Nordstrom black hole 
(hereafter referred to as `RN'), which represents
the groundstate of the simple
system studied in \cite{Liu:2009dm, Faulkner:2009wj}, 
has a `residual' zero-temperature entropy.
This degeneracy is exact in the classical $N\to\infty$ limit;
at finite $N$ one expects it to be lifted to a large low-lying density of states.
It is likely that the non Fermi liquid behavior does not depend
 on the large low-energy density of states: the small-frequency
behavior depended on the {\it existence} of the IR CFT, not on the large
central charge $c \propto s(T=0)$ of the IR CFT.

A closely related question regards the stability of the extremal black hole geometry.
It is a stable solution of the Einstein-Maxwell theory.  However,
many known $AdS$ string vacua
which UV-complete this model contain charged boson fields
which at finite density and low temperature will exhibit the holographic superconductor instability
\cite{Gubser:2008px, Hartnoll:2008vx}.
Conveniently, the physical systems to which we would like to
apply these models also
generically exhibit a superconducting instability (\eg \cite{highTc, heavyfermion}):
the $T=0$ limit of most known non-Fermi liquids
is under a superconducting dome\footnote
{
Other possible groundstates for holographic finite-density systems, 
for example resulting
from the presence of neutral bulk scalars,
have been explored recently in \cite{Herzog:2009gd, Gubser:2009qt, notaholographicmottinsulator}. 
}. 
The calculations in the RN black hole provide
a model for the ``normal" state above the superconducting $T_c$.

As discussed in the last section of \cite{Faulkner:2009wj}, this raises the following very
natural question:
what happens to the holographic
Fermi surface in the presence of superconductivity?
One might expect to see a gap in the spectral weight, and
we will see below that this is realized.
Unlike the fermion two-point function calculation,
here there are some choices for the bulk action.
In addition to choosing the self-couplings of the bulk scalar $\scalar$,
one must decide how to couple the scalars to the bulk spinor field $\spinor$.
It is always possible to include a $|\scalar|^2 \bar \spinor \spinor$ coupling.
In duals of matrix-like theories,
where the spinor field is dual to an operator of the form
$\tr \lambda $, it is
natural to include a scalar $\scalar$ with twice the charge of $\spinor$,
dual to the operator $ \tr \lambda \lambda $ \cite{shamit}.
Its dimension at strong coupling is not determined by this information.
In this case, a (as it turns out, much more interesting) coupling of the schematic form $\spinor \spinor \scalar^\star $ is permitted by gauge invariance.
We will specify the spinor structure of the coupling below.

The effect of this coupling is to pair up modes at the Fermi surface,
in a manner extremely similar to the Bogoliubov-deGennes
understanding of
charge excitations of a BCS superconductor.

Interestingly, if the mass-to-charge ratio of bulk scalars is large enough,
they do not condense \cite{Denef:2009tp},
and we pause here to comment on this case.
This in itself is an interesting phenomenon which does not happen at weak coupling, and
should be explored further.
 It means that the criteria for a string vacuum which exhibits the
Fermi surfaces described in \cite{Liu:2009dm, Faulkner:2009wj}, but {\it not} the superconducting
instability, are reminiscent of those required of a string vacuum
which could be that of our universe: one doesn't want light scalar fields.
In the latter context, a large machinery \cite{modulistabilization}
has been developed to meet the stated goal, and similar techniques will be useful
here.
In such a case, the calculation of \cite{Liu:2009dm, Faulkner:2009wj}
is valid to very low temperatures.
One effect which cuts this off is the following\footnote{
We thank Nabil Iqbal for an instructive conversation on this point.}.
In the RN black hole background, there is a finite density of
fermions in the bulk \cite{resistivitypaper}.
There is a Fermi surface
(in the sense that the bulk-to-bulk fermion spectral density has
a nonanalyticity at $\omega=0, k= k_F$).
There are interactions between these bulk fermions
mediated by fluctuations of the metric and gauge field.
The Coulomb force is naively always stronger \cite{weakgravity},
but can be screened.  This leaves behind the
interactions by gravity, which are universally attractive.
There is some similarity with phonons.
Of course, these interactions are suppressed by powers of $N^2$
(where $N^2 \equiv G_N^{-1}$ in units of the $AdS$ radius).
This may lead to BCS pairing with an energy scale
\be T_c \sim \varepsilon_F^{{\rm bulk}} e^{ - { 1\over \nu(0) V} } \sim \mu e^{ -N^2 } \ee
where $\nu(0)$ is the density of states at the bulk Fermi surface,
and $V\sim N^{-2}$ is the strength of the attractive interaction.
This is a very small temperature.
This is exactly the scale of the splitting between the degenerate groundstates
over which the RN black hole averages
which is to be expected at finite $N$.
Nevertheless,
this is one way in which the
RN black hole groundstate of the system studied in \cite{Liu:2009dm, Faulkner:2009wj}
is unstable, without the addition of  extra scalar degrees of freedom.

\bigskip

In this paper, we will probe
(a few examples of) holographic superconducting groundstates
with fermionic operators.
The retarded Green's functions $G_R(\omega, k)$ we compute may be compared
with data from angle-resolved photoemission experiments
on cuprate superconductors \cite{campuzano, zxshen}.
In these experiments, a high-energy photon knocks
an electron out of the sample, which is then detected.
Knowing the energy and momentum of the incident photon
and measuring the energy and momentum of the detected electron
allows one to infer that the sample has an electronic
excitation specified by their difference;
the intensity of the signal is proportional to the density of such states,
$A(\omega, k) \equiv \Im G_R(\omega, k)$
(at least in the sudden approximation, which is believed
to be valid for the relevant photon frequencies).
Actual photoemission experiments have the limitation
that they can only kick electrons out of {\it occupied} states,
and hence can only measure an intensity $I \propto A(\omega, k) f(\omega)$, where
$f(\omega)$ is the Fermi factor, which at zero temperature vanishes for $\omega $
above the chemical potential.
We do not have this limitation.

Lest the reader get the wrong idea, we emphasize here
several
features of our calculations which differ from the experimental situation
in any strongly-correlated electron system.
Perhaps most glaringly,
as in previous work,
the Fermi surfaces we discuss in this paper
are {\it round}.  There is no lattice in our system.
At short distances, our theory approaches a relativistic
conformal field theory; the UV conformal symmetry is
broken explicitly by finite chemical potential
(we will also comment on the effects of a small temperature).
Also, our superconducting order parameter has $s$-wave symmetry,
and so there are no nodes at which the gap goes to zero.
It would be very interesting to improve upon this situation.

Above the superconducting critical temperature $T_c$,
one usually has gapless excitations at $k = k_F$.
When one cools the superconductor below $T_c$,
the locus $\{ k = k_F\} $ generally remains the
{\it surface of minimum gap},
\ie\ the locus in momentum space where
the gap in the fermion spectral density is smallest.
This is not precisely the case here.
This is because in general
the holographic superconducting condensate also
affects the geometry outside the horizon region, \ie\ UV physics,
and changes the effective Schr\"odinger potential
determining value of $k$ at which the Dirac bound state occurs.
The difference between $k_F$ without the condensate
and the surface of minimum gap  will be small in the examples we study,
which have $T_c$ small compared to $\mu$,
and are therefore close to the RN geometry as we review below.

\cite{chenkaowen} appeared as this paper was being completed.
The crucial Majorana coupling is not included there.
Related work will appear in \cite{fabio, allan}.

\subsection{Why the Majorana coupling is important}
\label{sec:BdG}

We focus here on the case of odd $d$ 
(the number of spacetime dimensions of the boundary field theory),
where a single Dirac spinor in the bulk describes a single Dirac
spinor operator in the boundary theory.
In the case of even $d$, we will need to couple
together {\it two} bulk Dirac fields.

\def\CC{C}
The bulk action we consider for the fermion is
\bea
\label{majoranaaction}
S[\spinor] &=& \int d^{d+1}x
\sqrt{-g}
\[  i \bar \spinor \( \Gamma^M D_M  - \mspinor  \)\spinor
\right. \cr
&+& \left.\eta_5^\star \scalar^\star  \spinor^T  \CC  \Gamma^5 \spinor + \eta_5 \scalar \bar \spinor \CC \Gamma^5  \bar \spinor^T
\]~~.
\eea
$\scalar$ is the scalar field whose condensation spontaneous breaks
the $U(1)$ symmetry.
$\CC$ is the charge conjugation matrix,
which we specify below,
and $\Gamma^5$ is the chirality matrix, $ \{ \Gamma^5, \Gamma^M\} = 0$.
The derivative $D$ contains the coupling to both 
the spin connection and the gauge field
$ D_M \equiv \partial_M + {1\over 4} \omega_{MAB}\Gamma^{AB} - i q_\spinor A_M$.
We will occasionally refer to the coupling to the scalar in
\eqref{majoranaaction} as a `Majorana coupling'
because $ \spinor^T  \CC \Gamma^5 \spinor$ is like a Majorana mass term.
One reason for the necessity of
the antisymmetric charge conjugation operator between
the fermion fields in this term is that
the simpler-looking object
$ \spinor_\alpha \spinor_\alpha $ is zero because
the components are grassmann-valued.

As we will describe, the coupling
$\scalar^\star \spinor^T  \CC  \spinor + {\rm h.c.}$
is also possible, but does not accomplish the desired effect.
The coupling with the $\Gamma^5$ arises in descriptions
of fermionic excitations of color superconductors 
\cite{Alford:2007xm}.  In that context,
the chirality matrix is required by parity conservation;
since $\varphi$ there is a bilinear of the same quarks to which
it is coupling, the intrinsic parity of the quarks cancels out.

One could worry that the perturbations of the scalar field will mix
(in the sense that one will source the other)
with the fermion equations of motion.
This does not happen in the computation of two-point functions because of fermion
number conservation.

We pause here to note the instructive
similarity between \eqref{majoranaaction} and the action governing electrons
in a BCS $s$-wave superconductor
\bea
&&S[c] = \int d^{d-1}k d\omega
\( c^\dagger_\alpha (\omega, k) \( i  \omega - \xi_k \) c_\alpha(\omega,k)  \right.
\\&&\nonumber
\left.
- \Delta(k) c^\dagger_\uparrow(\omega,k) c^\dagger_\downarrow(-\omega,-k)
- \Delta^\star(k) c_\uparrow(\omega,k) c_\downarrow(-\omega,-k)
\)
\eea
where $\alpha = \uparrow, \downarrow$ are spin indices,
$\xi_k \equiv v_F(|\vec k|- k_F)$,
and $\omega$ is measured from the chemical potential.
This similarity is instructive because
it explains why other couplings between
the spinor and the condensate do not automatically produce a gap.

The basis of modes which diagonalizes such an action is
the Nambu-Gork'ov basis:
\be \gamma_\alpha(k) \equiv u(k) c_\alpha(k)+
\CC^\alpha_\beta v(k) c_\beta^\star(-k)  \ee
note that $u$ and $v$ do not have spin indices.
The Green's function which results from this mixing is
\be
\vev{c_k(\omega)^\dagger c_{k}(\omega)}_R
= { \omega + \xi_k\over \( \omega+i\epsilon \)^2 - \xi_k^2 - |\Delta(k)|^2 }.
\ee
This function has {\it two} poles for each $k$;
they approach $\Re(\omega) = 0 $ as $k$ approaches the Fermi surface.
Each has a minimum real part of order $\Delta$.
The residues of these two poles, however, varies with $k$:
at large negative $k-k_F$, the weight is mostly in the pole with
$\Re(\omega) < 0$ and the excitations is mostly a hole.  As $k$ moves through $k_F$,
the weight is transferred to the other pole, and the excitation is mostly an electron.
Without such a mixing between positive and negative frequencies,
the Green's function would have only one pole,
which would be forced to cross $\Re(\omega) = 0$
as $k$ goes from $k \ll k_F $ to $k \gg k_F$,
and there could not be a gap.
This continuity argument assumes that in the absence
of the condensate, the dispersion is monotonic.

\section{Review of groundstates of holographic superconductors}

Consider the action
\be
{\cal L} = {1\over \kappa^2} \( R + {6 \over L^2} - {1\over 4} (dA)^2 -
\left| ( \nabla - i \qscalar  A)\scalar \right|^2  - \mscalar^2 |\scalar|^2 \).
\ee
We will work in units where the AdS radius $L$ is unity.
For $\mscalar^2 -2\qscalar^2 < -3/2$, the Reissner-Nordstrom AdS solution is unstable at low temperature to forming scalar hair. The extremal limit of these hairy black holes was found in \cite{Horowitz:2009ij}\footnote
{
Groundstates of holographic superconductors,
including other forms of the scalar potential, were also studied in
\cite{Gubser:2009cg}.  Our analysis should apply to those
whose IR region is $AdS_4$; we leave the other cases for future work.
}. Unlike the extreme Reissner-Nordstrom black hole, the area of the horizon goes to zero in this limit. The detailed behavior near the horizon depends on $\mscalar$ and $\qscalar$, but for $\mscalar^2\le 0$, the solution has Poincare symmetry near the horizon. This has an important consequence. Consider solutions of the Dirac equation with $e^{ik_\mu x^\mu}$ dependence. If $k$ is timelike in the near horizon region, then one can impose the usual ingoing wave boundary condition to compute the retarded Green's function $G_R$. Since the boundary condition is complex, the Green's function is complex, and hence Im $G_R$ is nonzero indicating a continuous spectrum of states. However, if $k$ is spacelike, the solutions are exponentially growing or damped. Normalizablility requires the exponentially damped solution. This is a real boundary condition, and so the solutions will be real and Im $G_R = 0$.  This is qualitatively different from the extreme Reissner-Nordstrom AdS whose near horizon geometry is $AdS_2\times R^2$. In that case, there is a continuous spectrum for all $(\omega, k^i)$.

\begin{figure}[h!]
\begin{center}
\includegraphics[scale=.7]{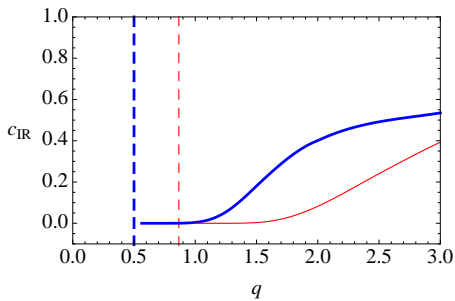}
\caption{
The speed of light in the IR CFT, $\cIR$, as a function of the boson charge.
The blue thick curve is $\mscalar^2=-1$,
the red thin curve is $\mscalar^2 = 0$.
The vertical dashed lines indicate the value of $\qscalar$ below which the RN solution
is stable.
 \label{fig:vs}}
\end{center}
\end{figure}

The light cone in the near horizon region will not have the same speed of light
as the asymptotic geometry. One can show that as one approaches $\mscalar^2 -2\qscalar^2 = -3/2$ where the RN solution becomes stable, the
speed of light in the IR CFT, $\cIR$
(
not to be confused with the  central charge of the infrared CFT), goes to zero (see FIG.~\ref{fig:vs}).
  This means that in momentum space, the light cone opens up so all momenta are effectively timelike, and the spectrum continuously matches onto the RNAdS case.

In more detail, the static, plane symmetric solutions take the form:
\be\label{eq:metric}
 ds^2=-g(r) e^{-\chi(r)} dt^2+{dr^2\over g(r)}+r^2(dx^2+dy^2)
\ee
\be
A=\phi(r)~dt, \quad \scalar = \scalar(r)~~.
\ee
For $\mscalar^2=0$, the zero temperature solution not only has Poincare symmetry but approaches $AdS_4$ near the horizon, and $r=0$ is just a Poincare horizon.  The leading order corrections can be found analytically and depend on a parameter $\alpha$ which is a function of $\qscalar$, but stays small ($ |\alpha |< .3$). Explicitly,
\bea
&&\phi = r^{2+\a},\quad  \scalar = \scalar_0 - \scalar_1 r^{2(1+\a)},
\cr\cr &&
\quad \chi =\chi_0 - \chi_1 r^{2(1+\a)},
\quad g=r^2(1 - g_1r^{2(1+\a)})
\eea
where
\be
\qscalar \scalar_0 = \left({\a^2 + 5\a + 6\over 2}\right)^{1/2}, \quad \chi_1 = {\a^2 + 5\a + 6\over 4(\a + 1)}e^{\chi_o}
\ee
\be
g_1 =  {\a + 2\over 4} e^{\chi_o}, \quad \scalar_1 = {\qscalar e^{\chi_o} \over 2(2\a^2 + 7\a +5)}\left({\a^2 + 5\a + 6\over 2}\right)^{1/2}
\ee
Although the curvature remains finite, derivatives of the curvature  diverge at $r=0$ unless $\a = 0$.
FIG. 2 shows the solution for $g(r)$ and $\phi(r)$ for a choice of $\qscalar$ which is close to the value $\sqrt{3}/2$ where Reissner-Nordstrom AdS is stable. One sees that $g$ dips down and has a local minimum at a  value  $r\approx 1$. As  $\qscalar \rightarrow \sqrt{3}/2$, $g$ vanishes at this local minimum which becomes the horizon of the extremal Reissner-Nordstrom AdS black hole.

For $\mscalar^2 < 0 $ ( and $\qscalar^2>-\mscalar^2/6$), the zero temperature solution near the horizon is
 \be\label{qansatzz}
 \scalar=2(-\log r)^{1/2},\quad ~g= (2\mscalar^2/3) r^2 \log  r,\quad   e^\chi=-K\log r
\ee
\be
~\phi=\phi_0r^\beta (-\log  r)^{1/2},
\ee
where
\be
\beta=-\frac{1}{2} + \frac{1}{2}\left( 1-\frac{48 \qscalar^2}{\mscalar^2}\right)^{1/2}
 \ee
 and $\phi_0$ is adjusted to satisfy the boundary condition at infinity. The near horizon metric is (after rescaling $t$)
 \be\label{nhmetric}
ds^2 = r^2(-dt^2  + dx_idx^i) + {3 dr^2\over 2 \mscalar^2 r^2 \log r}
\ee
One clearly sees the Poincare symmetry (but not the conformal symmetry) in this case.
There is a rather mild null curvature singularity at $r=0$.

\begin{figure}[h!]
\begin{center}
\includegraphics[scale=.7]{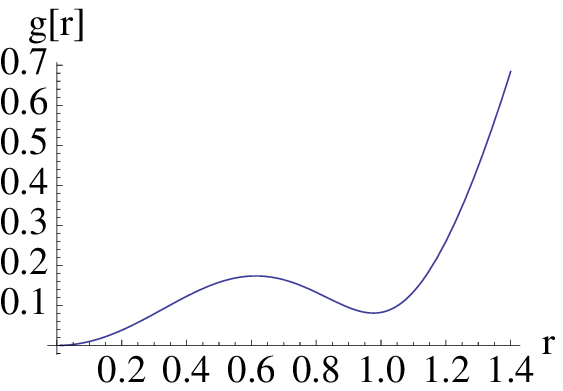}
\includegraphics[scale=.7]{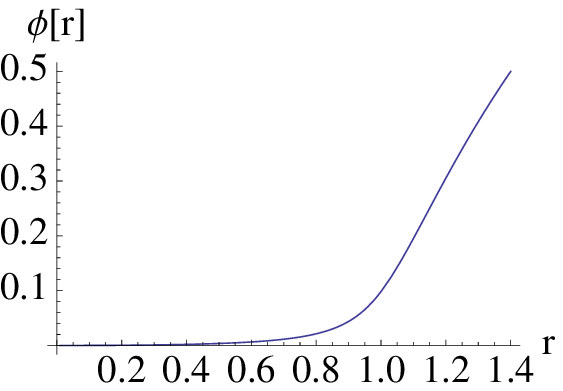}
\caption{
This plot of the emblackening factor $g$ (left)
and the electrostatic potential $\phi$ (right)
in the $\qscalar = 1.3, \mscalar^2 = 0$ groundstate solution
exhibits the almost-RN horizon at $r=1$.
In this plot and those below, we use units where $\mu = 2 \sqrt 3$.
 \label{fig:nearRN}}
\end{center}
\end{figure}

\section{Dirac equation}
\label{sec:de}


The Dirac action is
\begin{equation}
S_0 = i \int d^{d+1} x \,\sqrt{-g}  \, \overline{\spinor} \left( \Gamma^M D_M -  \mspinor -  \lambda | \scalar |^2  \right) \spinor
\end{equation}
where we are using the conventions of \cite{joebook}. The $\lambda$ coupling
could be replaced by a more general function  of $|\scalar|^2$. We
will set $\lambda = 0$ for now except to discuss its effects briefly below.

As discussed in section \ref{sec:BdG} if the charge of the
scalar is such that $\qscalar = 2 \qspinor$ then we can add to this
\begin{equation}
\label{etacoupling}
S_\eta = \int d^{d+1}x \, \sqrt{-g} \, \scalar^* \overline{\spinor^c} \left( \eta^* + \eta_5^* \Gamma^5 \right) \spinor + \mathrm{h.c}
\end{equation}
where the charge conjugation matrix is
\begin{equation}
\spinor^c = \CC \Gamma^{\ut} \spinor^* \qquad \left( \CC \Gamma^{\ut} \right) \Gamma^{\umu}
\left( \CC \Gamma^{\ut} \right)^{-1} = \Gamma^{\umu*}
\end{equation}
This term is essentially a Majorana mass term. There are two terms because
there are two Majorana spinors in the bulk (or Weyl spinors) and these can have independent
masses.

In the case of odd numbers of bulk dimensions, there is no $\Gamma^5$
and this term does not exist.
This matches the fact that in odd numbers of bulk dimensions,
a single Dirac spinor in the bulk describes
a {\it chiral} fermion operator in the boundary theory
\cite{Iqbal:2009fd};
such a fermion cannot be paired with itself in a rotation-invariant way.
The analogous coupling in odd bulk dimensions requires
two Dirac fermions.
That this is possible can be seen by dimensionally reducing
a theory with an even number of bulk dimensions on a circle.
We will not discuss this in detail here.

Now we study the Dirac equation in more detail.
It turns out that the same
simplification that appeared in the RN background occurs
for the more general metric (\ref{eq:metric}).
Very briefly, the form of the spin connection
\begin{equation}
\omega_{\hat t \hat r} =d t \, \sqrt{g^{rr} }\partial_r \left( \sqrt{g_{tt}} \right)
\quad \omega_{\hat i \hat r} = - d x^i \sqrt{g^{rr}}  \quad \ldots
\end{equation}
implies that
\begin{equation}
\frac{1}{4} \omega_{ab M} e^M_c \Gamma^c \Gamma^{ab}
= \frac{1}{4} \Gamma^r \partial_r \ln \left( - g g^{rr} \right)
\end{equation}
so we can rescale $\cF = ( - g g^{rr} )^{1/4} \spinor$ and remove the spin connection
completely. The new action is
\begin{equation}
S_0 = i \int d^{d+1} x\,  \sqrt{g_{rr} }\, \overline{\cF} \left( \Gamma^M D_M' -  \mspinor \right) \cF
\end{equation}
where $D_M' = \partial_M- i \qspinor A_M$ with no appearance of the spin connection.

The Dirac equation following from $S_0 + S_\eta$ is
\begin{equation}
\left( \Dslash' - \mspinor \right) \cF + 2 i \scalar (\eta - \eta_5 \Gamma^5 ) \CC \Gamma^{\ut} \cF^* = 0~.
\end{equation}
Expand this into Fourier modes with $k_x = k, k_y=0$:
\begin{equation}
\left( \Dslash'(k,\omega) - \mspinor \right) \cF(k,\omega) + 2 i \scalar (\eta - \eta_5 \Gamma^5 ) \CC \Gamma^{\ut}  \cF^*(-k,-\omega) = 0
\end{equation}
To get any further we must specify a basis of Dirac matrices. We
focus on
$d=3$, that is, a $3+1$ dimensional bulk.
We choose a basis of bulk Gamma matrices as in \cite{Faulkner:2009wj},
\bea
\Gamma^{\ur} &=& \begin{pmatrix} - \sigma^3 & 0 \\ 0 & - \sigma^3 \end{pmatrix}
\quad
\Gamma^{\ut} = \begin{pmatrix} i \sigma^1 & 0 \\ 0 & i \sigma^1 \end{pmatrix}
\quad
\Gamma^{\ux} = \begin{pmatrix} - \sigma^2 & 0 \\ 0 &  \sigma^2 \end{pmatrix}
\cr
&&\Gamma^{\uy} = \begin{pmatrix} 0 & \sigma^2 \\ \sigma^2 & 0 \end{pmatrix}
\quad
\Gamma^{5} = \begin{pmatrix} 0 & i \sigma^2 \\ -i \sigma^2 & 0 \end{pmatrix}
\eea
such that $\Gamma^{\ut *}  = - \Gamma^{\ut}$ and $\Gamma^{\ur*} = \Gamma^{\ur}$
which fixes the charge conjugation matrix to be $C \Gamma^{\ut} = \Gamma^{\ur}$.
This basis has the features that (with $\eta_5=0$ and $k_y=0$) the Dirac equation  is completely real.

We will now split the 4-component spinors into two 2-component spinors
$\cF = ( \cF_1, \cF_2 )^T$ where the index $\alpha = 1,2$ is the Dirac index
of the boundary theory. Then
\bea
0=\left( D_r( \pm k) + \sqrt{g^{tt}} \sigma^1  \omega \right)
\cF_{1,2}(k,\omega)
&&\cr 
- 2 i \sigma^3 \scalar \eta
\cF_{1,2}^*(-k,-\omega) \pm 2 i \sigma^1 \scalar \eta_5 \cF_{2,1}^*(-k,-\omega)
&&
\label{eq:above}
\eea
where\footnote
{
The frequency $\omega$ is measured from the chemical potential.
}
\begin{equation}
D_r(k) \equiv -\sqrt{g^{rr}} \sigma^3 \partial_r  - \mspinor - \sqrt{g^{xx}} i \sigma^2 k + \sqrt{g^{tt}}
\sigma^1 \qspinor A_t
\end{equation}
We see that the $\eta_5$ term mixes $\cF_1(k,\omega)$ with $\cF^*_2(-k,-\omega)$
(and $\cF_2(k,\omega)$ with $\cF^*_1(-k,-\omega)$) - this is the mixing that will most interest us, because for the RN background these two fields have coincident Fermi surfaces (at $\omega =0$). 
Setting
$\eta = 0 $ (\ref{eq:above}) becomes
\begin{equation}
\label{eq:thedir}
\left( D_r(\pm k) \otimes \mathbf{1} + \sigma^1 \otimes \begin{pmatrix}
\sqrt{g^{tt}} \omega  & \pm 2 i \scalar \eta_5 \\
\pm 2 i \scalar^* \eta_5^*   & - \sqrt{g^{tt}} \omega \end{pmatrix}
\right)
\Psi_{1,2} =0
\end{equation}
where
\be
\Psi_1 \equiv  \begin{pmatrix} \cF_1(k,\omega) \\ \cF_2^*(-k,-\omega) \end{pmatrix}
\qquad \Psi_2 \equiv  \begin{pmatrix} \cF_2(k,\omega) \\ \cF_1^*(-k,-\omega) \end{pmatrix}
 ~.
 \ee
 are the bulk analogs of the Nambu-Gork'ov spinor.\footnote{ The index on $\Psi_\alpha$
is the boundary theory Dirac index. For the rest of this section (\ref{sec:de})
for simplicity of the discussion we will concentrate mostly on one of these: $\Psi_1$. In section (\ref{sec:results})
we will give results for $\Psi_2$. }
We see explicitly from (\ref{eq:thedir}) that for a general black hole
background in the absence
of mixing ($\eta_5=0$) the spectrum of $\cF_1(k,\omega)$ 
compared to the spectrum of $\cF^*_2(-k,-\omega)$ is a reflection
about the $\omega=0$ axis. This is crucial for \emph{generically} generating
gapped states for non-zero $\eta_5$.
 
We have set $\eta=0$ both to make the analysis easier
and because turning on both $\eta$ and $\eta_5$
implies that some discrete symmetry of the boundary theory is broken.

\bigskip
For completeness, we record
the Dirac equation with $\eta_5 =0, \eta \neq 0$.
In this case, the mixing would be between
$\cF_1(k,\omega)$ and $\cF^*_1(-k,-\omega)$ with the equation being
\begin{equation}
\left( \begin{pmatrix} D_r(\pm k) & 0 \\ 0 & D_r(\mp k) \end{pmatrix} + \begin{pmatrix}
\sqrt{g^{tt}} \omega \sigma^1 & - 2 i \scalar \eta \sigma^3 \\
2 i \scalar^* \eta^* \sigma^3  & - \sqrt{g^{tt}} \omega \sigma^1 \end{pmatrix}
\right)
\tilde{\Psi}_{1,2} =0
\end{equation}
where $\tilde{\Psi}_{1,2} = ( \cF_{1,2}(k,\omega), \cF^*_{1,2}(-k,-\omega))^T$.
Because the differential operators $D_r$ in the diagonal entries above are evaluated
with opposite $k$, the two mixed components will not have coincident spectra at
$\omega = 0, \eta =0$
(see Figure 5 of \cite{Faulkner:2009wj} to see this in the RN background).
As such there will only be eigenvalue repulsion if there is some accidental eigenvalue
crossing, and this will generically occur away from $\omega=0$. 
This should be contrasted with the $\eta_5$ mixing discussed above.

\subsection{Boundary conditions}

As reviewed in section 2,  many of the solutions found in \cite{Horowitz:2009ij} have an emergent Poincare symmetry in the deep IR, and some even have emergent conformal symmetry.  For now
we will mainly consider the latter case in which the solution approaches $AdS_4$ near the horizon.
To determine the IR boundary conditions
for the spinor
appropriate for the retarded Green's function,
we consider the Dirac equation
in the far IR region,
where the metric is just pure $AdS_4$ with
no electric field and zero chemical potential:
\begin{eqnarray}
\label{IRAdS4}
\nonumber ds^2 &=&  r^2 \left( - \cIR^2 dt^2 + d\vec{x}^2 \right) + \frac{ L_{IR}^2 dr^2}{r^2} \\
\phi &=& 0 \quad \scalar = \scalar_0 \quad \chi = \chi_0~.
\end{eqnarray}
The speed of light in the dual IR CFT
is $\cIR = e^{-\chi_0/2}/L_{IR}$. The most relevant terms
in the Dirac equation
close to the Poincare horizon are $ \partial_r \Psi_1 = M \Psi_1$, with
\begin{equation}
\label{eq:nh}
M \equiv \begin{pmatrix}
 \frac{L_{IR}}{r^2} \left(  i \sigma^2 {\omega\over \cIR} - \sigma^1 k \right)
  & 0 \\
0 &  \frac{L_{IR}}{r^2} \left(  - i \sigma^2 {\omega\over \cIR} -  \sigma^1 k \right) \end{pmatrix}~.
\end{equation}
Very generally, the off-diagonal terms are subdominant, by arguments
given in \cite{Horowitz:2009ij} in the discussion of the Schr\"odinger potential for the optical conductivity: the relative magnitude of the off-diagonal term
to the terms appearing in (\ref{eq:nh}) is $\scalar \sqrt{g_{tt}} = \scalar \sqrt{g} e^{-\chi/2}$ which must
generally vanish on the horizon.

Because of the diagonal form of \eqref{eq:nh},
we can
construct a basis of ingoing solutions by
considering $\cF_1(k,\omega)$ and $\cF_2^*(-k,-\omega)$ separately.
As is familiar from zero-temperature $AdS$, the
character of the boundary conditions depends on the sign of $s^2 \equiv -\omega^2/\cIR^2 + k^2$.
To begin with we will work outside the light cone where $s^2 > 0$ is spacelike. Here the behavior of
the solutions is normalizable and non-normalizable. We will pick the one
which is normalizable at $r\to0$:
\bea
\label{eq:norm}
&&(\bI) ~~
 \cF^{*\, \bI}_2(-k,-\omega) \buildrel{r\to0}\over{\approx}
 0,~~~
\cF^{\bI}_1(k,\omega)
\buildrel{r\to0}\over{\approx}
 \xi_{N}^{\bI} e^{ - s L_{IR} /r}
= \cr\cr
&&\begin{pmatrix} \sqrt{ k + \omega/\cIR} \\ - \sqrt{k - \omega/\cIR } \end{pmatrix}
\exp\left( - \sqrt{k^2 - {\omega^2\over\cIR^2 }} {L_{IR} \over r} \right);
\eea
$\xi_N^{\bI}$ is an eigenvector of the matrix $M$ in \eqref{eq:nh}.
This now allows
us to formulate the general incoming boundary conditions in order
to compute retarded correlators. We simply use the $i\epsilon$ prescription
to define how to continue the branch cuts in (\ref{eq:norm}) to timelike $s^2 < 0$. That is,
we take $\omega \rightarrow
\omega + i \epsilon$.

For the other component $\cF_2(-k,-\omega)$ we can simply take $\omega \rightarrow - \omega$
in (\ref{eq:norm}),
\bea
\label{eq:norm2}
&&(\bII) ~~
\cF^{\bII}_1(k,\omega) \buildrel{r\to0}\over{\approx}
 0,~~
\cF^{*\,\bII}_2(-k,-\omega)
\buildrel{r\to0}\over{\approx}
 \xi_{N}^{\bII} e^{ - s L_{IR} /r} = \cr\cr
&&\begin{pmatrix} \sqrt{ k - \omega/\cIR} \\ - \sqrt{k + \omega/\cIR } \end{pmatrix}
\exp\left( - \sqrt{k^2 - \frac{\omega^2}{\cIR^2} } \frac{L_{IR} }{r} \right) ,
\eea
and again for timelike $s^2 <0$ in (\ref{eq:norm2}) we continue $\omega \rightarrow \omega + i \epsilon.$\footnote
{Beware the following confusion: because there is a complex conjugation
on $\cF_2^*$, one might expect this to switch the sign of the $i \epsilon$.
This is not
the case because we should think of analytically continuing $\cF_2^*(-k,-\omega)
\rightarrow \cF_2^*(-k,-\omega^*)$; this procedure preserves the incoming
boundary conditions.}
In the absence of the $\eta_5$ mixing, the two solutions of the Dirac equation
\eqref{eq:thedir} for $\Psi_1$
specified by the IR behavior $\bI, \bII$
compute the Green's functions for the two
boundary fermion operators,
 $\CO_1(\omega,k)$ and $\CO^\dagger_2(-\omega,-k)$.

Now we consider the boundary conditions at the boundary of the UV $AdS_4$. This
will tell us how to read off the field theory correlators.
The mixing term is again subdominant at the UV boundary, so that the asymptotic
behavior is the same as usual:
\begin{equation}
\cF^{\bI}_1(k,\omega) \buildrel{r\to\infty}\over{\approx} \begin{pmatrix} B^{\bI}_1 r^{-\mspinor } \\ A^{\bI}_1 r^{\mspinor } \end{pmatrix}
\quad
\cF^{*\, \bI}_2 (-k,-\omega) \buildrel{r\to\infty}\over{\approx} \begin{pmatrix} B^{*\, \bI}_2 r^{-\mspinor } \\ A^{*\, \bI}_2 r^{\mspinor } \end{pmatrix}~~
\end{equation}
and similarly for $\bI \to \bII$.
The boundary retarded Green's function is: \footnote{The minus
signs appearing in front of $A_2^{*\, \bI, \bII}$ come from the fact
that $- (A_2^*)^\dagger $ is the source for $\mathcal{O}^\dagger_2$
where the minus sign is from anti-commuting this (Grassman valued) source in the boundary
theory action so that it is in the correct order and the action is real.}
\begin{equation}
\label{eq:amatrix}
\begin{pmatrix} B_1^{\bI} &   B_1^{\bII} \\
 B_2^{*\, \bI} & B_2^{*\, \bII}  \end{pmatrix}
= \begin{pmatrix} G_{\cO_1 \cO^\dagger_1} &   G_{\cO_1 \cO_2} \\
G_{\cO_2^\dagger \cO_1^\dagger} & G_{\cO_2^\dagger \cO_2} \end{pmatrix}
\begin{pmatrix} A_1^{\bI} & A_1^{\bII} \\
 - A_2^{*\, \bI}  & - A_2^{*\, \bII}  \end{pmatrix}
\end{equation}
The definition of the Green's functions appearing above is:
\begin{equation}
G_{CD}(\omega,k) = i \int d^{d-1}x dt e^{i k x - i \omega t} \theta(t) \left< \left\{ C(x,t), D(0,0) \right\}
\right>
\end{equation}
Note that the spectral densities (which should be positive by unitarity)
are $\Im G_{\cO^\dagger_1 \cO_1}$ and $ \Im G_{\cO_2 \cO_2^\dagger}$.

More generally including the analysis for $\Psi_2$ the above matrix
(\ref{eq:amatrix}) will fit into the Lorentz covariant correlator which is
a $4\times 4$ matrix (recall that this is for $k_x = k, k_y = 0$):
\begin{equation}
\begin{pmatrix}
G_{\cO \cO^\dagger} &   G_{\cO \cO_{c}^\dagger} \\
G_{\cO_{c} \cO^\dagger} & G_{\cO_{c} \cO^\dagger_{c}}
\end{pmatrix} = \begin{pmatrix}  G_{\cO_1 \cO_1^\dagger} & 0 & 0 & G_{\cO_1 \cO_2} \\
0 & G_{\cO_2 \cO_2^\dagger} & G_{\cO_2 \cO_1} & 0 \\
0 & G_{\cO_1^\dagger \cO_2^\dagger} & G_{\cO_1^\dagger \cO_1} & 0 \\
G_{\cO_2^\dagger \cO_1^\dagger} & 0 & 0 & G_{\cO_2^\dagger \cO_2}
\end{pmatrix}
\end{equation}
where $\cO = (\cO_1, \cO_2)^T$ and $\cO_{c} = (C\gamma^t) (\cO^\dagger)^T$ where the
boundary theory charge conjugation matrix can be shown to be $C\gamma^t = 1$.
Note that all the entries in this $4\times 4$ matrix will be non-zero 
if both $\eta, \eta_5$ are turned on.

\subsection{Evolution equation}

It turns out there is a super nice way to package the above linear differential
equation into a non-linear evolution equation, in the spirit of the evolution
equations considered in \cite{Iqbal:2008by} and used in
\cite{Iqbal:2009fd, Liu:2009dm}.
This is useful for identifying the Fermi surfaces numerically,
because although the spinor components vary greatly with $r$,
their ratios, which appear in the evolution equation, remain order one.

Define the following matrices:
\bea
Y \equiv \begin{pmatrix} \left(\cF^{\bI}_1\right)_1 &  \left(\cF^{\bII}_1\right)_1 \\
  \left(\cF^{*\bI}_2\right)_1 &  \left(\cF^{*\bII}_2\right)_1 \end{pmatrix} && 
Z \equiv \begin{pmatrix} \left(\cF^{\bI}_1\right)_2 &  \left(\cF^{\bII}_1\right)_2 \\
-  \left(\cF^{*\bI}_2\right)_2 & - \left(\cF^{*\bII}_2\right)_2 \end{pmatrix}
  \cr\cr
 G &\equiv& Y Z^{-1}
\eea

Then one can write the following evolution equation:

\bea
\label{eq:evol}
&&\left( \sqrt{g^{rr}} \partial_r + 2 \mspinor \right) G =
 \\ \nonumber&&
G \left(  \sqrt{g^{xx}} k \sigma^3 + \sqrt{g^{tt}} (\omega  +
\qspinor A_t \sigma^3 )
+ 2 i \scalar \begin{pmatrix}  0& \eta_5 \\ - \eta_5^* & 0 \end{pmatrix} \right) G
\\ \nonumber&&+ \left(  - \sqrt{g^{xx}} k \sigma^3 + \sqrt{g^{tt}} (\omega +
\qspinor A_t  \sigma^3) + 2 i \scalar \begin{pmatrix}  0& -\eta_5 \\ \eta_5^* & 0 \end{pmatrix} \right)
\eea

The boundary conditions on this matrix at the IR $AdS_4$ horizon and the UV $AdS_4$ boundary become respectively:
\bea
G  &\buildrel{r \to 0}\over{\approx}&  \begin{pmatrix} - \sqrt{ \frac{ k+ \omega/\cIR}{k - \omega/\cIR} } & 0 \\
0 & \sqrt{\frac{ k- \omega/\cIR}{k + \omega/\cIR} } \end{pmatrix}
\cr
G  &\buildrel{r \to \infty}\over{\approx}&  r^{- 2 \mspinor }  \begin{pmatrix} G_{\cO_1 \cO^\dagger_1} &   G_{\cO_1 \cO_2} \\
G_{\cO_2^\dagger \cO_1^\dagger} & G_{\cO_2^\dagger \cO_2} \end{pmatrix}
\eea
Note that when $\eta_5 = 0$ the evolution equation preserves the diagonal
form of the initial condition in the IR.

This method runs into difficulty if $Z$ becomes non-invertible
at finite $r$; this happens for the multi-node boundstates 
associated with secondary Fermi surfaces.


\section{Results: Bound states outside the emergent light cone}
\label{sec:results}

\subsection{No mixing}

We start by looking at $\eta_5 = \eta = 0$ so there is no mixing. We will
concentrate on the field $\cF_2(k,\omega)$ (from which we
can reflect about $\omega =0$ to generate  $\cF_1^*(-k,-\omega)$. )
Note that we are now switching $1\leftrightarrow 2$ relative to the discussion of the
previous section - all results in this section will be for the Nambu Gork'ov spinor $\Psi_2$.
The reason being the primary Fermi surface in the RN background 
(the one with largest $k_F$)
makes its appearance in the Green's function for 
$\cF_2(k,\omega)$ \cite{Faulkner:2009wj}. We are interested in understanding the
fate of this primary Fermi surface in the condensed phase.

Now since the initial conditions are real for spacelike $s^2 > 0$ and the Dirac equation for $\cF_2$
in the absence of mixing is real, the spectral functions should be zero outside the
emergent IR lightcone. This is true up to delta functions which can appear because
the real part of the Green's function has a pole which becomes a delta
function in the imaginary part thanks to Kramers-Kronig.
These are bound states of the Dirac equation
since they are normalizable at the IR $AdS_4$ horizon and at the UV $AdS_4$ boundary.
They will represent infinitely long lived fermion states in the field theory.

For now we will look for these states in a small set of the zero temperature hairy black holes constructed in \cite{Horowitz:2009ij} and reviewed above.
We will concentrate on the case with zero scalar potential energy
($V(\scalar) =0 \rightarrow \mscalar^2 =0$)
with general charge $\qscalar$ for the scalar. In this case $L_{IR} = 1$ and the
speed of light in the IR CFT can be found numerically, see FIG.~\ref{fig:vs}.

The fermion charge\footnote{
There is a factor of two difference in the normalization of the charges for both scalars and spinors in
\cite{Liu:2009dm}
(LMV) compared to \cite{Horowitz:2009ij} (HR) -- they are related by $q_{LMV} = 2 q_{HR}$. We will work with the $q_{HR}$ normalization throughout.} will be constrained by gauge invariance
to be $\qspinor = \qscalar/2$, so that
the $\eta_5$ term can be added later.
The mass of the fermion is a priori independent
of the mass of the scalar.
We work with $\mspinor = 0$ for numerical convenience. It will
be interesting to look at small charges close to the critical charge $\qscalar  \rightarrow \sqrt{3}/2$ where the critical temperature $T_c \rightarrow 0$ and $\cIR \rightarrow 0$. For $\qscalar < \sqrt{3}/2$ the RN black hole
is stable, and as can be seen from FIG.~\ref{fig:nearRN},
the superconducting groundstate approaches the RN solution.
In this limit the spectral densities should look more and more similar to the ones
of the RN black hole, which we have a good handle on. Indeed, for reference,
we know that there is a Fermi surface in the RN black hole for $\mspinor = 0$ and $\qspinor = \sqrt{3}/4$ when $k_F \approx 0.75$ with IR scaling exponent $\nu \approx .18$.

\begin{figure}[h!]
\begin{center}
\includegraphics[scale=.7]{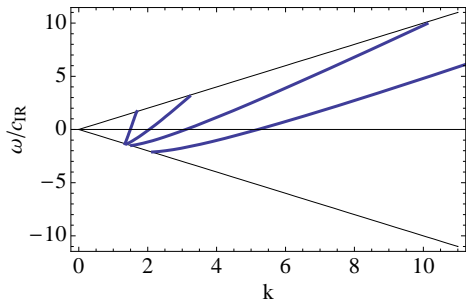}
\caption{Boundstates outside the IR lightcone
for various values of $\qscalar = 1.5, 2, 3, 5$. Note that the frequency
axis has been scaled by $c_{IR}$ which is different for each curve.
We can't resist mentioning the
approximate relation $k_F \approx \qscalar$ for where the curves
cross $\omega=0$.
\label{fig:boundstates}}
\end{center}
\end{figure}

FIG.~\ref{fig:boundstates} shows the location
of these states for different $\qscalar$ in these 
zero temperature superconducting
backgrounds. 

\subsection{Mixing}

The stable gapless ($\omega=0$) excitations we have found in FIG. 3 seem rather surprising in a strongly coupled theory. We now demonstrate that turning on $\eta_5 \neq 0$ (and keeping $\eta = 0$)  the stable excitations studied above develop a gap.
 The reason
for this can be simply understood by the general arguments of eigenvalue
repulsion. Since the positive frequency modes mix with the negative frequency
modes (at the \emph{same} $k$) the repulsion occurs when $\omega = 0$.

More carefully, we can study the Dirac equation with mixing.
Because the initial conditions (\ref{eq:norm}) and (\ref{eq:norm2}) for spacelike $s^2 < 0$
are real one might expect that again the spectral functions are zero except
for delta functions. This is a little
subtle because the Dirac equation (\ref{eq:thedir}) is real except for the $\eta_5$
term.  However it  turns out that despite this, the spectral functions are still
zero. We can see this in two ways. Firstly, the spectral functions are the difference
in the retarded and advanced Green's functions (this is more general
than the \emph{imaginary part} of the retarded function). For spacelike $s^2 >0$ these
two Green's functions are calculated with the same Dirac equation and the \emph{same} initial conditions (the difference comes from the $i \epsilon$ prescription when going to $s^2 < 0$.) Hence $G_R = G_A$ here and the spectral function is zero except for on bound states.

Secondly,  the evolution equation (\ref{eq:evol}) for spacelike $s^2 >0$ preserves
the following form of the $2\times 2$ Green's function matrix $G$ (recall we have
switched $1\leftrightarrow 2$ relative to (\ref{eq:evol}) ): 
\be 
G_{\cO_2 \cO_2^\dagger } , \,  
G_{\cO_1 \cO_1^\dagger} \in \mathbb{R} \qquad G_{\cO_2 \cO_1}, \, (G_{\cO_1^\dagger \cO_2^\dagger})^* \in e^{i\arg(\eta_5)}\mathbb{R}~.
\ee 
Hence the spectral densities for $G_{\cO_2 \cO_2^\dagger }, 
G_{\cO_1 \cO_1^\dagger}$ are zero. The phase of $\eta_5$ is arbitrary since we can change
it by rephasing the operator $\mathcal{O}$, hence it cannot matter for the spectral
density of $G_{\cO_2 \cO_1}$.

\begin{figure}[h!]
\begin{center}
\includegraphics[scale=.265]{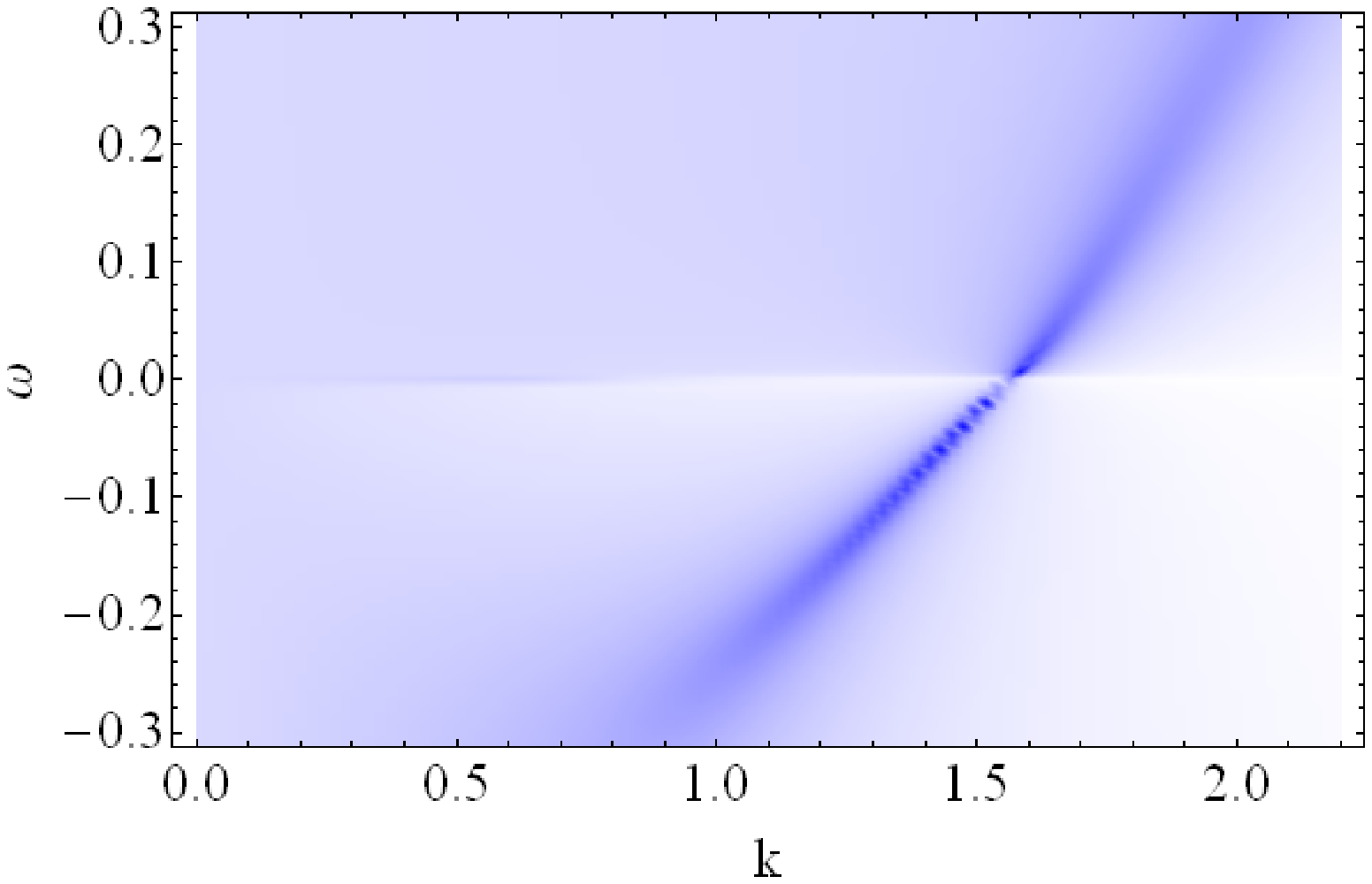}
\includegraphics[scale=.265]{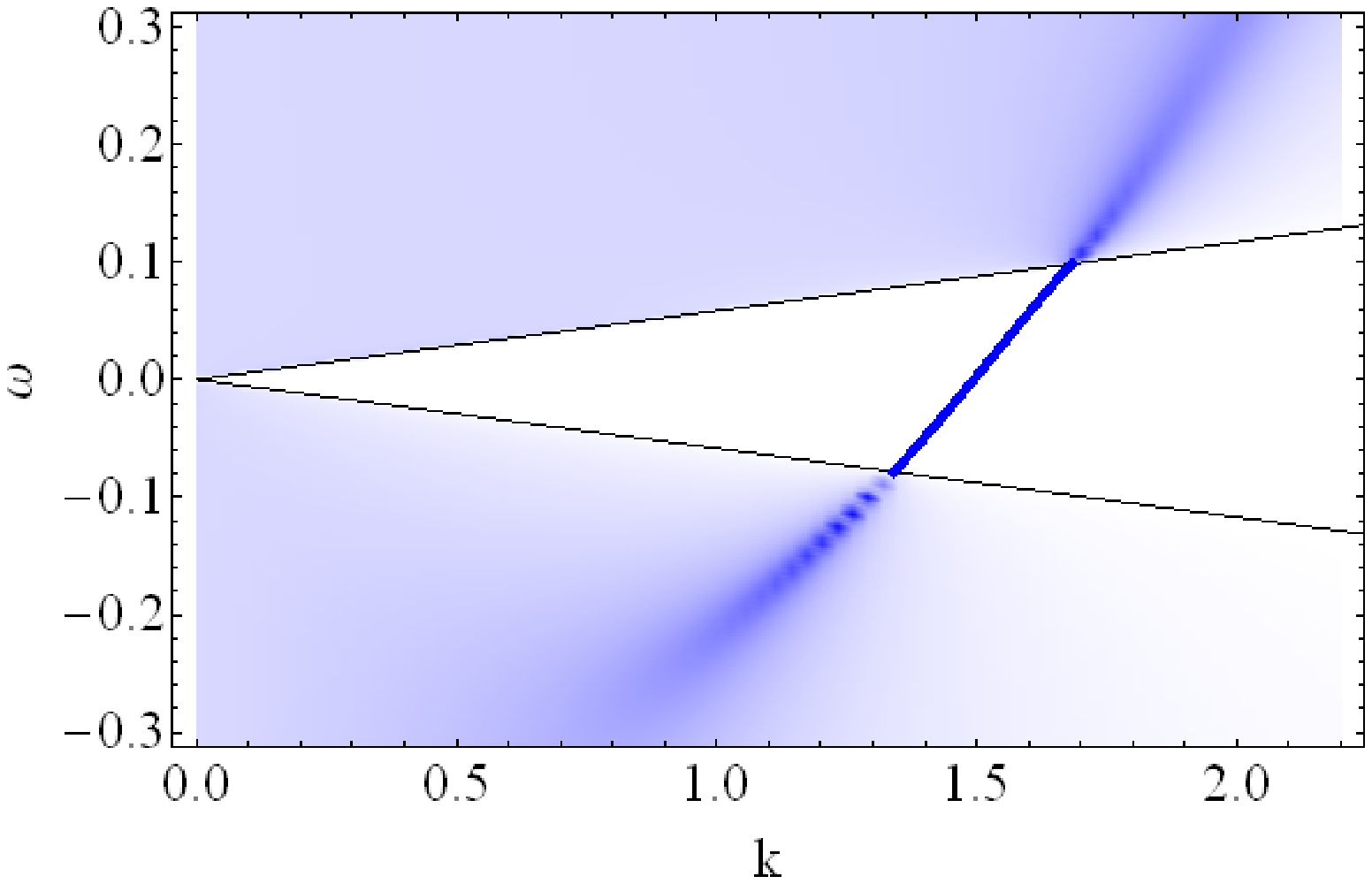}
\includegraphics[scale=.265]{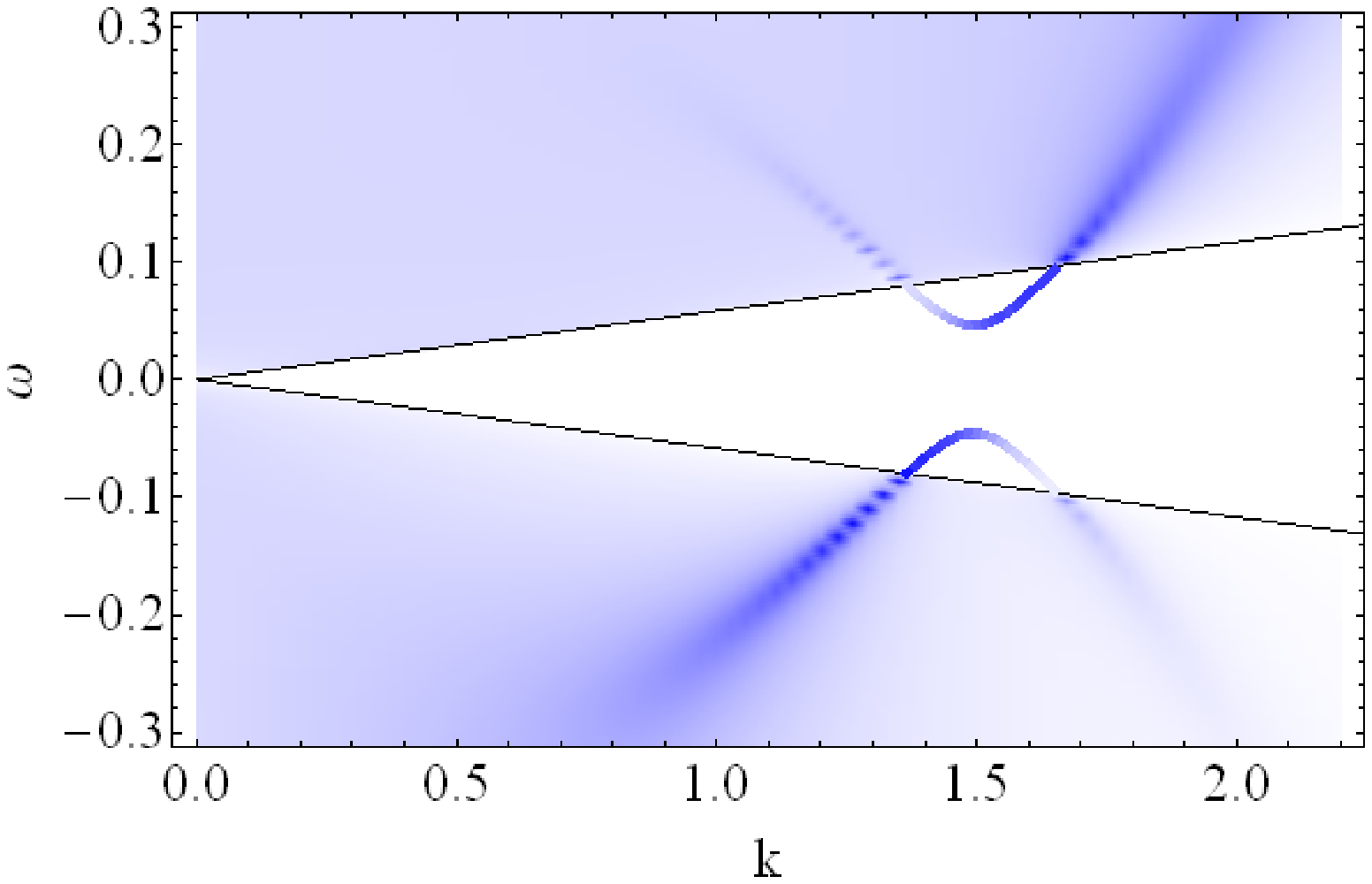}
\includegraphics[scale=.265]{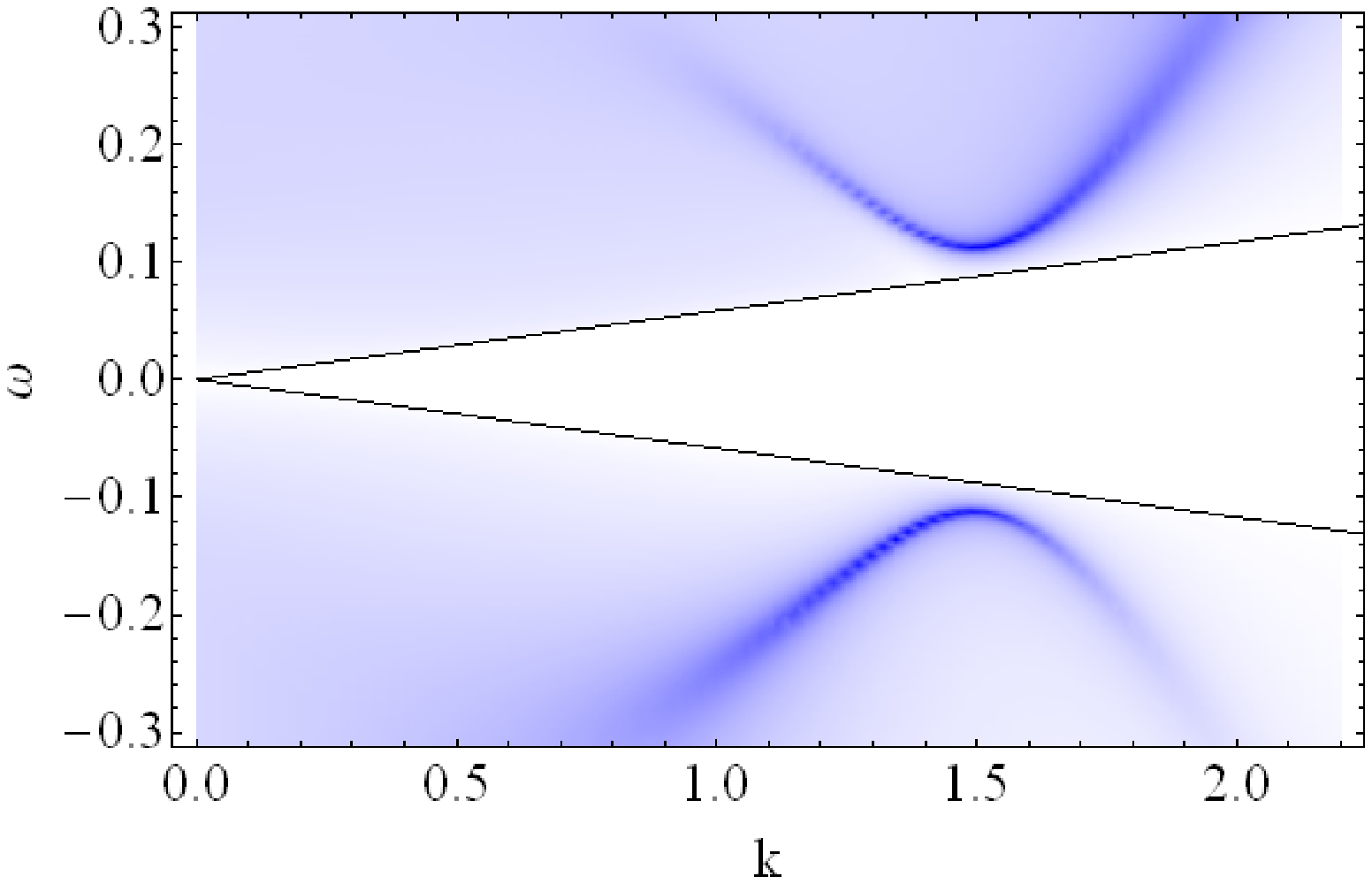}
\caption{
Mixing between positive- and negative-frequency modes
due to the Majorana coupling.
Shown are density plots of the fermion spectral density $A(k, \omega)
= {\rm Im} G_{\cO_2 \cO_2^\dagger}$
for $\qspinor = {3 \over 4}, \mspinor = 0 $.
The first plot is in the $T=0$ RN black hole, no scalar.
The remaining plots are
in the zero temperature background with
$ \qscalar = {3\over 2}, \mscalar^2 =0 $,
for various values of the Majorana coupling,
$\eta_5 = 0, 0.2, 1.5 $.
 \label{fig:gap} }
\end{center}
\end{figure}

\begin{figure}[h!]
\begin{center}
\includegraphics[scale=.8]{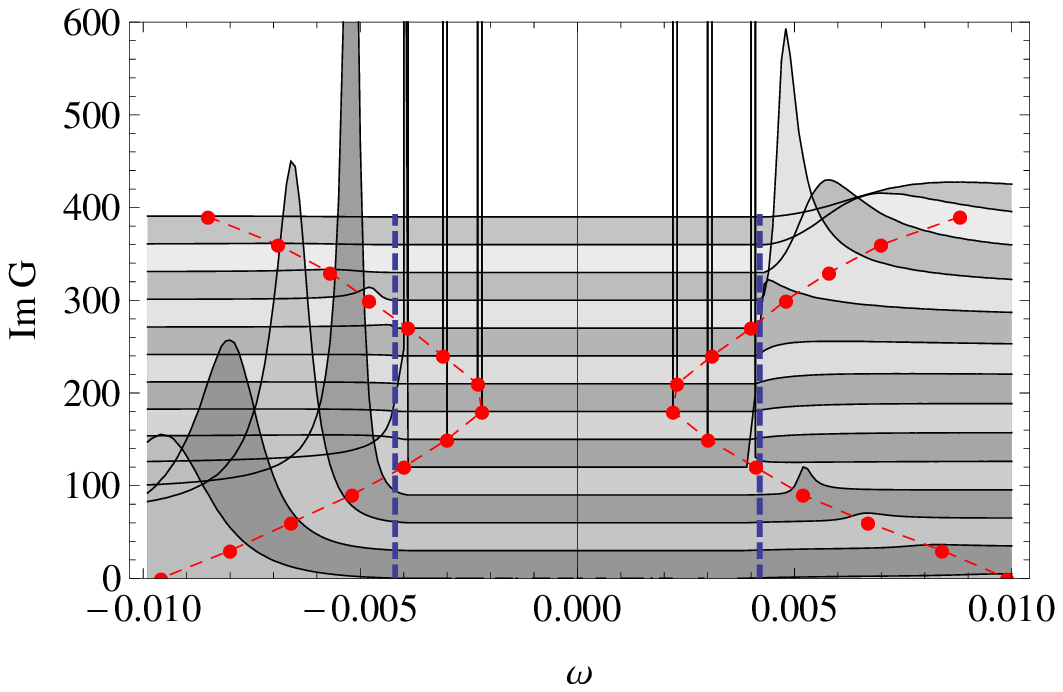}
\includegraphics[scale=.8]{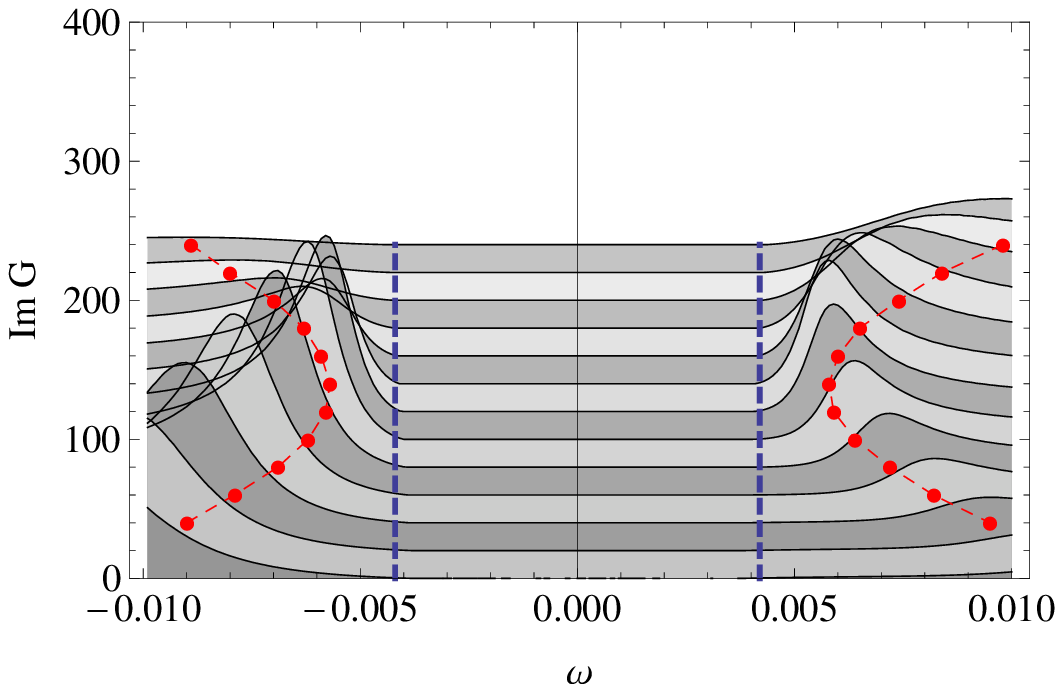}
\caption{ The effect of the Majorana coupling on the fermion spectral density. Shown are plots
of $A(k, \omega)$ at various $k \in [.81, .93]$ for $\qspinor = \half, \mspinor = 0 $ in a
low-temperature background of a scalar with $ \qscalar = 1, \mscalar^2 =-1$, with $\eta_5 = 0.025$ (top)
and $\eta_5 = 0.075$ (bottom). The blue dashed line indicates the boundary of the region in
which the incoherent part of the spectral density is completely suppressed, and the lifetime of
the quasiparticle is infinite. The red dotted line indicates the location of the peak.
 \label{fig:gap3}}
\end{center}
\end{figure}

To find the bound state in this situation we should look for places where
$\det G^{-1} = 0$ at the boundary. Note that $\det G^{-1} \in \mathbb{R}$ for $s^2 >0$
so indeed this is a well defined problem. This delta function will appear
in all $4$ spectral densities. The residue however will be different in each component.
We concentrate on 
$G_{\cO_2 \cO_2^\dagger}$ because this is what should be
accessible to photoemission ``experiments''. The results are given in  FIG.~\ref{fig:gap} and FIG.~\ref{fig:gap3}.


\def\ccF{\chi}
We can learn something from perturbation theory in $\eta_5$.
The splitting is determined by
the eigenvalues of the matrix
\be
V \equiv \( \begin{matrix}
P_\uparrow & Q_\uparrow\cr
Q_\downarrow & P_\downarrow
\end{matrix}
\)
\ee
where
\be P_\alpha
\equiv \int dr \sqrt{g_{rr}} \bar \ccF_\alpha^{(0)} \omega \sqrt{g^{tt}} \ccF_\alpha^{(0)} (-1)^\alpha
=\omega J^t_{\alpha\alpha}
\ee
($J$ was defined in \cite{Faulkner:2009wj}, appendix C)
and
\be Q_\uparrow
\equiv \int dr \sqrt{g_{rr}} \bar \ccF_\uparrow^{(0)} 2 i \eta_5 \scalar  \ccF_\downarrow^{(0)},
Q_\downarrow
\equiv \int dr \sqrt{g_{rr}} \bar \ccF_\downarrow^{(0)} 2 i \eta_5^\star \scalar^\star  \ccF_\uparrow^{(0)}
\ee
where
$ \ccF_\alpha^{(0)} $ denotes the boundstate wavefunction
in the basis
$ \ccF_\uparrow = \cF_1, \ccF_\downarrow = \cF^\star_2(-\omega,-k).$
Thinking of the Dirac equation as a Schr\"odinger problem,
this matrix $V$ is the perturbation Hamiltonian in the
degenerate subspace.

The fact that at $\omega=0, \eta_5 =0$, the up and down boundstates are the same
implies that
$ P_\uparrow = - P_\downarrow \equiv P$ and $Q_\uparrow = Q_\downarrow^\star$);
the eigenvalues of $V$ are therefore
\be  \pm \sqrt{-P^2 + |Q|^2 } .\ee
Looking for low-energy boundstates
with fixed $k$ then
requires these eigenvalues to vanish,
which occurs when $ -P^2 + |Q|^2 = 0 $,
\ie\ when $ \omega \sim |\eta_5|$.

\subsection{Luttinger-like behavior near the lightcone}

To understand what's happening at $ \omega^2 = \cIR^2 k^2$,
we consider the Schr\"odinger form of the wave equation,
where the role of the energy eigenvalue is played by $-k^2$.
For simplicity (and because the pictures are nicer) we draw the potentials for the case of a
charged scalar probe (not to be confused with the charged
scalar $\scalar$ which is condensing.)  For further details, see Appendix B of \cite{Faulkner:2009wj}.

\begin{figure}[h!]
\begin{center}
\includegraphics[scale=.5]{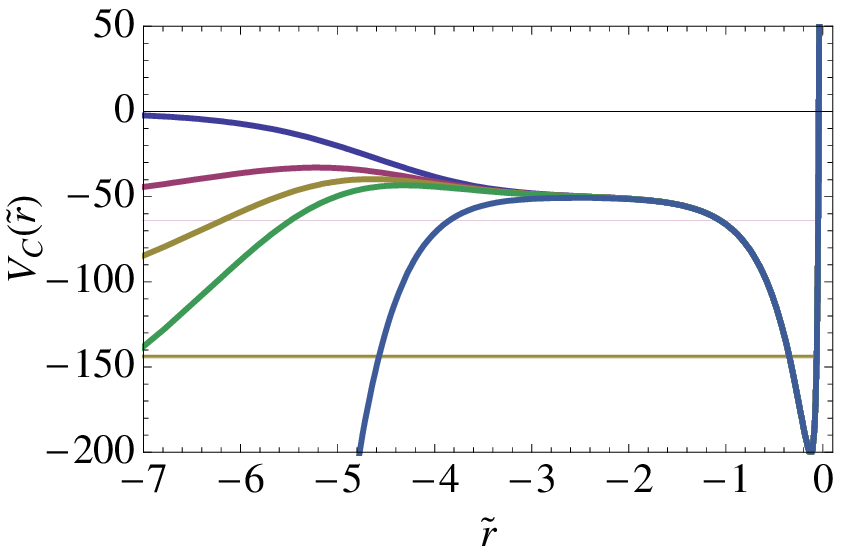}
\includegraphics[scale=.5]{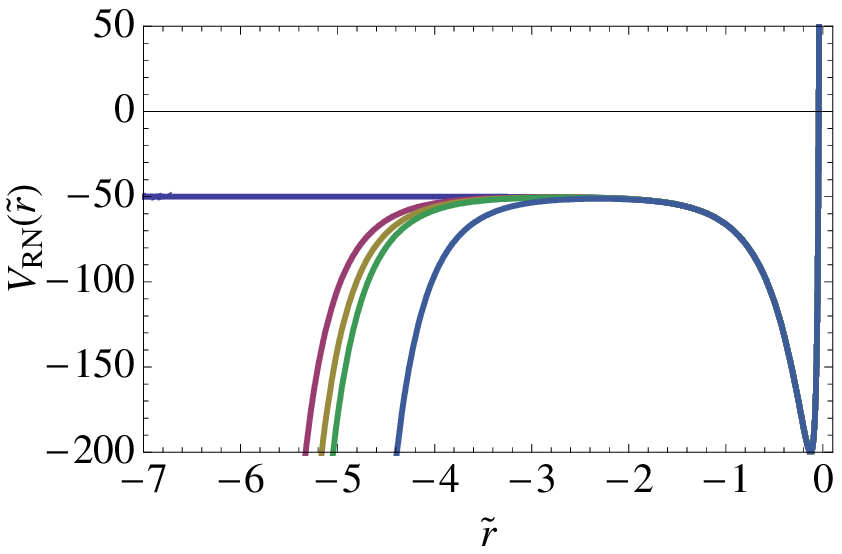}
\caption{ The effective Schr\"odinger potentials for a probe \emph{scalar} with mass $m_{\rm
probe}^2 = -3/2$ and $q_{\rm probe} = 5$. The horizontal axis is the tortoise coordinate; the UV
boundary is to the right, at $\tilde{r}=0$. The different curves are different values of $\omega > 0$;
the top curve in each plot is for $\omega=0$.
Left: The potentials for the groundstate with $\mscalar^2 = 0$ and $\qscalar = 1$. Right: The corresponding pictures for the RN black hole with the
same charge density. \label{fig:schrod}}
\end{center}
\end{figure}

The physics of the IR lightcone is visible in FIG.~\ref{fig:schrod}. In the RN background (right
plot), turning on any nonzero frequency opens up a bottomless pit in the effective potential
leading into the $AdS_2$ region where the tortoise coordinate $\tilde{r} \to - \infty$. Therefore, in
the RN groundstate there are no infinitely-stable quasiparticles with nonzero frequency. On the
other hand, in the superconducting groundstate, the limiting value of the effective potential as
$\tilde{r} \rightarrow - \infty$ is $- \omega^2/c_{IR}^2$.
Therefore, there is a threshold frequency $|\omega| = |\cIR k| $
below which the IR limit of the Schr\"odinger potential
remains above the boundstate energy.
More precisely,
there will be a normalizable bound state close to the boundary as long as the energy ($-k^2$)
is less than the limiting value $- \omega^2/c_{IR}^2$.
Beyond this the bound state enters the light-cone and is no longer
a stable quasi particle.

The fact that we see a stable particle below the continuum
is qualitatively what one expects for systems with a gap $\omega_0$.  For energies $\omega_0 < \omega <2\omega_0$, one excites a single quasiparticle which is stable since there is nothing for it to decay into. Only at energies above $2\omega_0$ does one start to see a continuum.

\bigskip

The spectral density near the lightcone, and in particular the width of the quasiparticle after
it enters the lightcone can be computed by matching between the $AdS_4$ regions in UV and IR as
in \cite{Faulkner:2009wj, Horowitz:2009ij}. The size of the overlap region is controlled by the
quantity $s^2 = k^2-\omega^2/\cIR^2$ which should be small in units
of the chemical potential. In the notation of \cite{Faulkner:2009wj}, the result for
the Green's function is of the form
\be
\label{eq:matching} G \sim \left( B_+ +  B_- \sG \right) \left(A_+ + A_- \sG \right)^{-1}
\ee
where $A_{\pm}, B_\pm$ are {\it real}\footnote{They are only real if we take $\eta_5 \in i
\mathbb{R}$ which we can do without loss of generality.} data associated with the UV region, and
$\sG$ is the IR CFT Green's function to be discussed below. If there is mixing between positive
and negative frequency modes then $A_{\pm}, B_{\pm}$ are $2\times 2$ matrices
in the basis of the Nambu-Gork'ov spinor. They are smooth
(analytic) functions of $k,\omega$ so the leading non analytic behavior in $k,\omega$ is from
$\sG$.   For purposes of exposition we will describe the results for a probe scalar field in
parallel to that of the spinor. We will leave details of the spinor calculation to Appendix A.

For a probe scalar, the IR CFT Green's function is
\be
\label{scalarsG} \sG \sim \begin{pmatrix} \( k^2 - {\omega^2 \over \cIR^2} \)^{\nu_c^+}  & 0 \\
0 & \( k^2 - {\omega^2 \over \cIR^2} \)^{\nu_c^-}
\end{pmatrix} .
\ee
The quantities $\nu_c^\pm$ are related to the IR CFT scaling dimension of the 
boundary operator by
$\Delta_{IR}^\pm = {d\over 2 } + \nu_c^\pm$, 
and are determined by studying the behavior of the field at
the UV boundary of the IR $AdS_4$ region in \eqref{IRAdS4}. They are given by
\be \nu_c^\pm \equiv \sqrt{ \({d \over 2}\)^2 + L_{IR}^2(m_{\rm probe}^2 \pm  |\eta_5|\scalar_0 ) } ,\ee
where $\scalar_0 = \scalar(r=0)$ (the subscript $c$ is for `condensed' and is intended to
distinguish this object from the analogous IR CFT scaling dimension in the $AdS_2$ region of RN
\cite{Faulkner:2009wj}). Notice that the IR CFT scaling dimension {\it depends on the coupling
$\eta_5$}. 

For the probe spinor the IR CFT Green's function appearing in
(\ref{eq:matching}) is
\begin{equation}
\label{sping} \sG \sim \begin{pmatrix}  \sqrt{
\frac{k+\omega/\cIR}{k-\omega/\cIR} }  & 0  \\
0  &  \sqrt{ \frac{k-\omega/\cIR}{k+\omega/\cIR} } \end{pmatrix} \( k^2 -\frac{\omega^2}{\cIR^2}\)^{\nu_c}
\end{equation}

For the spinor case the relation between $\Delta_{IR}$ and $\eta_5$ is,
\be \nu_c \equiv  L_{IR} \sqrt{\mspinor^2 +4 |\scalar_0 \eta_5|^2 } \qquad
\Delta_{IR} = d/2 + \nu_c  ,\ee see Appendix A for more details.

We can extract two interesting statements from these calculations. From the form of
\eqref{eq:matching} (and in particular the reality of $A,B$) we learn that at generic $\omega,
k$ (but small $|s|$ so that this matching applies),
\be \Im G \propto (B_- - B_+ A_+^{-1} A_- ) \left( \Im \sG \right) A_+^{-1}
\  ~.
\ee
The dependence of $\nu_c$ on $\eta_5$
has the following consequence. In the last plot of FIG.~\ref{fig:gap}, one can see that the
coupling to the condensate is also suppressing the incoherent spectral weight inside the
lightcone. This is because the IR CFT dimension is becoming large as we make $\eta_5$ large.

Finally, if we look near a quasiparticle pole, which close to the light-cone occurs when  $\det
A_+ = 0$, we see that the imaginary part of the location of the pole is determined by the IR CFT
Green's function. This determines the width of the resonance as it enters the lightcone. The
result is that the width behaves as
\be
\label{eq:wid}
\Gamma \sim (\omega- \cIR k)^{\nu_c^\pm}  \qquad\Gamma \sim (\omega- \cIR k)^{\nu_c \pm 1/2}
\ee
for the scalar and spinor respectively, which can be compared to the behavior in
FIG.~\ref{fig:gap}.\footnote{ Actually we need to be more careful for the case
$\nu_c < 1/2$ (for the spinor.) Here (\ref{eq:wid}) should be replaced by,
 $ \Gamma \sim (\omega- \cIR k)^{1/2 \pm \nu_c}$. }


We emphasize that there are two mechanisms which suppress the spectral weight: one sets it
exactly zero (except for delta functions) outside the light cone. This is a property of IR
behavior of the background geometry. The other mechanism suppresses the weight independently of
the momentum
 (this is a numerical observation visible from the dashed blue lines in
 FIG.~\ref{fig:gap3}), and depends on the scalar-spinor coupling.
This mechanism generates the gap for the quasiparticle peak, and can be understood in terms of
the dependence of the effective IR scaling dimension of the fermion operator on $\eta_5$ as in
the previous discussion. The latter mechanism also affects physics at $k=0$ whereas the
lightcone mechanism does not.

%


\section{Discussion}

We should make a few remarks about the effects of
other possible couplings between the bulk spinor and scalar.
The coupling
\be
\label{neutralaction}
S_{{\rm neutral}}[\spinor] = - i \int d^{d+1}x
\sqrt{-g}
\lambda |\scalar|^2 \bar \spinor\spinor
\ee
is possible whatever the charge of the spinor and scalar.
By the argument given in section \ref{sec:BdG},
the Green's function 
for the system with $\eta_5 = 0$
near $k_F$ should have only one pole
(whose location may however be dramatically affected by the couplings $\lambda, \eta$), 
and the effects of the interaction \eqref{neutralaction}
cannot be interpreted as mixing of particle and hole states.
As the $|\scalar|^2 \spinor^2$ coupling is varied,
it is easy to be fooled into thinking that
there is a gap even when there is not,
when looking at energy distribution curves
because the Fermi momentum moves with $\lambda$.

As observed first in \cite{fabio} increasing the mass
of the effective field in the IR (which can be achieved
by either including the above $\lambda$ coupling or changing the
UV mass: $m_{IR} = \mspinor + \lambda \scalar_0^2$) can lead to 
poles which never reach the $\omega=0$ axis and may be interpreted
as gapped. This mechanism for removing low-energy spectral weight
(which happens because increasing $m_{IR}$
pushes up the effective Schr\"odinger potential for the 
Dirac equation)
is qualitatively different from the mixing described in the previous sections. It is analogous
to adding a relativistic mass to particles and anti-particles in a relativistic
field theory at non-zero chemical potential. 
The gap in this situation is around $\omega = -\mu$ and does not generically
produce a gap at $\omega = 0$. This should be compared to
the gap from the $\eta_5$ coupling 
which is like adding a mass to particles and holes (absence of particles)
about the Fermi surface at $\omega =0$.

It was shown in \cite{Horowitz:2009ij} that the zero temperature superconductors we have studied here do not have a hard gap in the optical conductivity: The real part of the  conductivity remains nonzero (although typically exponentially small) at low frequency and $T=0$.
Despite this fact, one might have wondered whether such a hard gap
in the conductivity
exists for the fermionic probes that we study in this paper.
The existence of a non-zero spectral weight around the origin
of FIG.~\ref{fig:gap} suggests that this is not the case; however, to see this effect in the conductivity it would
be necessary compute a $1/N^2$ correction as in \cite{resistivitypaper}.

It would be interesting to understand better what property
of the boundary theory
is reflected by
the presence of the $\eta_5$ coupling, which
is required to produce an actual gap in the fermion response.
One clue is that its presence specifies
the `intrinsic parity' of the dual operator,
\ie\ the dual operator acquires an
interesting phase under a parity transformation.
Realizing string vacua where this coupling
is nonzero would probably be valuable.

So far we have considered the fermion spectral function at zero temperature. FIG 7 shows what happens as one raises the temperature. The temperatures shown are much less than $T_c$. As $T\rightarrow T_c$, the condensate goes to zero, so its coupling to the fermions goes to zero and the gap disappears. Actually, the thermal broadening of the peak makes the gap disappear at about $.7T_c$. In the opposite limit, as $T\rightarrow 0$, the width of the peak vanishes rapidly. It appears to vanish faster than a power law, but the general temperature dependence deserves further investigation.

\begin{figure}[h!]
\begin{center}
\includegraphics[scale=.75]{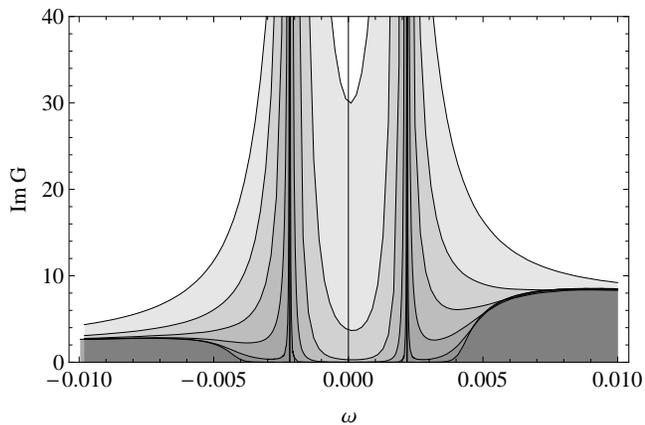}
\caption{The effect of temperature (much less than $T_c$) on the
fermion spectral function.
Shown are plots at $ \qscalar = 1, \mscalar^2 =-1,  \qspinor = \half, \mspinor = 0  , \eta_5 = .025$, and momenta where the peak is closest to $\omega =0$. The different curves correspond to different temperatures approaching $T=0$.  \label{fig:temp}}
\end{center}
\end{figure}

We close with a few comparisons with real phenomena.
Here we make a simple observation
which follows from the sharpness of the peaks in the `no man's land'  regime
(\ie\ outside the IR light cone).
This regime is induced by the superconducting order.
This means that if we start at high temperature in the normal phase
with some Fermi surface {\it without} stable quasiparticles (like say
a marginal Fermi liquid case, $\nu=\half$ in the notation of \cite{Faulkner:2009wj}),
and cool into the superconducting phase,
sharp quasiparticle peaks appear,
at least for $\eta_5$ not too big.
This matches a mysterious piece of cuprate phenomenology:
in the normal phase, photoemission experiments show no stable
quasiparticle peak, but a coherent peak
emerges in the superconducting phase
(see \eg figure 47 of the review \cite{damascelli}).
From the gravity point of view, this is
happening because
the scalar condensate is
removing the $AdS_2$ region
which was responsible for the finite lifetime of the
holographic quasiparticles \cite{Faulkner:2009wj}:
this is the gravity statement that
the condensate is lifting the many gapless excitations
into which the quasiparticle could decay.
The mechanism for the stability of these excitations
is very similar
to the recent
holographic explanation
\cite{Gubser:2009qf}
of the critical velocity in a (holographic) superfluid below which there is no drag,
and above which energy is dissipated by the creation of IR $AdS_4$ unparticles.

This similarity can be made more precise.
In a BCS super{\it{fluid}}, 
the decay of the quasiparticles
can be mediated by emission of a Goldstone
boson (this mode is eaten in a superconductor, and the 
following effect is absent).
It can happen that this 
decay is kinematically forbidden:
the decay cannot happen
if the group velocity of the quasiparticle
is larger than the speed of sound (see appendix B of \cite{zwerger}).
In our system, the quasiparticles
develop a finite lifetime when they can decay
into the modes of the IR CFT dual
to the IR $AdS_4$ region.
These modes are distinct from the Goldstone mode
(which is apparently hidden by powers of $N$),
but the effect is the same.


The energy distribution curves ($A(k,\omega)$ at fixed $k$)
shown in FIG.~\ref{fig:gap3}
exhibit another feature in common with ARPES measurements on the
cuprates, namely the so-called `peak-dip-hump' structure:
in addition to the quasiparticle peak, one sees a broad maximum
at larger $\omega$.
This is a consequence of the IR lightcone.
Over-ambitiously,
if this were the correct interpretation,
the location of the hump
would give a measurement
of the speed of light of the quantum critical theory.

\vspace{0.2in}   \centerline{\bf{Acknowledgements}} \vspace{0.2in}
We thank Nabil Iqbal and Hong Liu for collaboration on related matters. We thank 
S.~Hartnoll, S.~Kachru,
A.~Ludwig, M.~Mulligan, Y.~Nishida, S.~Sachdev, T.~Senthil, B.~Swingle, A.~Yarom, W.~Zwerger and many of the
participants of the ``Quantum Criticality and the AdS/CFT Correspondence" miniprogram at the
KITP for useful discussions. Work supported in part by funds provided by the U.S. Department of
Energy (D.O.E.) under cooperative research agreement DE-FG0205ER41360. The research was
supported in part by the National Science Foundation under Grant No. NSF PHY05-51164 and the
UCSB Physics Department. G. H. and M. R. were supported in part by NSF grant PHY-0855415.


\appendix

\section{Spinor in the IR $AdS_4$ region}

The Dirac equation in the IR $AdS_4$ region including the mixing term is
\bea
\nonumber && \left(
\begin{pmatrix}
\partial_r + \sigma^3 L_{IR} \mspinor/r & 2 i \sigma^2 \scalar_0 \eta_5 L_{IR} /r \\
2 i \sigma^2  \scalar^*_0 \eta_5^* L_{IR}/r & \partial_r + \sigma^3 L_{IR} \mspinor/r
 \end{pmatrix}
 +
\right.
 \cr
 &&\left.
\frac{L_{IR}}{r^2} \begin{pmatrix} k \sigma^1 - i \sigma^2 \omega/\cIR & 0 \\
 0 & k \sigma^1 + i \sigma^2 \omega/\cIR \end{pmatrix} \right) \Psi   = 0
 \eea

Now  to solve this we employ the following basis rotation $\cF^*_2 \rightarrow \sigma^1
\cF^*_2$. Then the Dirac equation takes the form:
\bea&&
\left( \partial_r +
 \frac{L_{IR}}{r^2} ( k \sigma^1 - i \sigma^2 \omega/\cIR ) \otimes 1
 \right. \cr
  &&\left.
+ \frac{L_{IR} }{r} \sigma^3 \otimes \begin{pmatrix}  \mspinor &   2i \scalar_0 \eta_5 \\  -2i
\scalar_0 \eta_5^* & -  \mspinor  \end{pmatrix}  \right) \bar{\Psi} = 0
\eea
where
\be
\bar{\Psi} = \begin{pmatrix} \cF_1(k,\omega) \\ \sigma^1 \cF^*_2(-k,-\omega) \end{pmatrix}~~.
\ee
We can now block diagonalize this equation into two independent Dirac equations. We make the
following basis rotation:
\begin{equation}
U  \begin{pmatrix} \mspinor L_{IR}  &  2 i \eta_5 \scalar_0  L_{IR}  \\  -2 i \eta_5^* \scalar^*_0
L_{IR} & -  \mspinor L_{IR}   \end{pmatrix}  U^{-1} =
  \begin{pmatrix}  - \nu_c &  0 \\ 0 & + \nu_c  \end{pmatrix} ~~.
\ee
Here,
\bea&&
 \nu_c = L_{IR} \sqrt{\mspinor ^2 +4 |\scalar_0 \eta_5|^2 }
 \cr \cr
 &&
  U = \begin{pmatrix} \mspinor L_{IR} -\nu_c & \mspinor L_{IR} +\nu_c \\
  - 2 i \eta_5^* \scalar^*_0 L_{IR} &  - 2 i \eta_5^* \scalar^*_0  L_{IR} \end{pmatrix}~~~.
\eea
where $\nu_c$ determines the conformal dimension of the spinor in the IR $AdS_4$ region. These
Dirac equations are then exactly that of a spinor in $AdS_4$ with mass $\pm \nu_c/L_{IR}$. The (two)
general incoming solutions can be found, and at the boundary of this IR $AdS_4$, a basis for
these solutions behaves like
\be
 \begin{pmatrix} \cF_1^{\bI} (k,\omega) \\ \sigma^1 \cF^{*\bI}_2(-k,-\omega) \end{pmatrix}
  \sim 1 \otimes U \begin{pmatrix} r^{\nu_c} \\ \mathcal{G}_{IR}(k,\omega) r^{-{\nu_c}} \\ 0 \\ 0 \end{pmatrix}
\ee
\be
  \begin{pmatrix} \cF_1^{\bII} (k,\omega) \\ \sigma^1 \cF^{*\bII}_2(-k,-\omega) \end{pmatrix}
  \sim 1 \otimes U \begin{pmatrix} 0 \\ 0 \\ \mathcal{G}_{IR}(k,-\omega) r^{-\nu_c} \\ r^{\nu_c}  \end{pmatrix}
\ee
where the IR Green's function for a spinor is
\begin{equation}
\mathcal{G}_{IR} (k,\omega) \sim \frac{\Gamma(1/2-\nu_c)}{\Gamma(1/2+\nu_c)} \sqrt{
\frac{k+\omega/\cIR}{k-\omega/\cIR} } \( k^2 -\frac{\omega^2}{\cIR^2}\)^{\nu_c}~~.
\end{equation}

We can then integrate these solutions out to the UV boundary where we can use similar methods to
(\cite{Faulkner:2009wj}) to read off a general form for the full Green's function. The result is
(\ref{eq:matching}).


\begin{thebibliography}{99}
\bibitem{shamit}
This argument for the genericity of
charge-$2q_F$ bosons is due to Shamit Kachru.
\end{thebibliography}
\end{document}